%% file: ms.tex
\PassOptionsToPackage{hyphens}{url}

\documentclass[sigconf, nonacm]{acmart}
\usepackage{xcolor}
\usepackage{multirow}
\usepackage{graphicx} 
\usepackage{hyperref}
\usepackage{xspace} 
\usepackage[hyphens]{url}  
\newcommand\rt{Reverse Address Translation\xspace}
\usepackage{hyperref}
\AtBeginDocument{%
  }

\begin{document}

\title{Analyzing Reverse Address Translation Overheads in Multi-GPU Scale-Up Pods}

\author{Amel Fatima}
\email{af3szr@virginia.edu}
\affiliation{%
  \institution{University of Virginia}
  \city{Charlottesville}
  \state{VA}
  \country{USA}
}

\author{Tuan Ta}
\email{tuan.ta@amd.com}
\affiliation{%
  \institution{AMD Research}
  \city{Bellevue}
  \state{WA}
  \country{USA}
}

\author{Bradford M. Beckmann}
\email{brad.beckmann@amd.com}
\affiliation{%
  \institution{AMD Research}
  \city{Bellevue}
  \state{WA}
  \country{USA}
}

\renewcommand{\shortauthors}{Fatima, Ta, and Beckmann}
\begin{abstract}
Distributed ML workloads rely heavily on collective communication across multi-GPU, multi-node systems. Emerging scale-up fabrics, such as NVLink and UALink, enable direct memory access across nodes but introduce a critical destination-side translation step: translating Network Physical Addresses (NPAs) to System Physical Addresses (SPAs), which we term Reverse Translation (\rt). Despite its importance, the performance impact of \rt remains poorly understood. In this work, we present the first systematic study of \rt in large-scale GPU clusters. Using an extended \textit{ASTRA-sim} framework with Omnet++ as the network backend, we model Link MMUs and Link TLBs and evaluate their effect on All-to-All collective communication across varying input sizes and GPU counts. Our analysis shows that cold TLB misses dominate latency for small, latency-sensitive collectives, causing up to 1.4× performance degradation, while larger collectives benefit from warmed caches and experience diminishing returns from over sized TLBs. Based on these observations, we propose two avenues for optimization: fused pre-translation kernels that overlap \rt with computation and software-guided TLB prefetching to proactively populate likely-needed entries. These techniques aim to hide translation latency, particularly for small collectives, improving throughput and scalability for inference workloads. Our study establishes a foundation for designing efficient destination-side translation mechanisms in large-scale multi-GPU systems.


\end{abstract}




\maketitle
\input{sec/1_intro}

\input{sec/2_background}
\input{sec/3_methodology}

\input{sec/4_key_ideas}

\input{sec/5_summary}
\input{sec/7_conclusion}
\bibliographystyle{ACM-Reference-Format}
\bibliography{bib/misc,bib/translation,bib/multigpu}
\appendix
\end{document}

%% file: sec/1_intro.tex
\section{Introduction}
Machine learning (ML) continues transforming daily life, powering AI-generated news summaries~\cite{concise2023}, assistant-like code generation tools~\cite{microsoftCopilot2023}, image synthesis~\cite{FacesinWild23_Borji}, and personalized recommendations~\cite{SurveyRS22_Ko}. At the heart of these applications lie large-scale models—particularly language models—which have ballooned from millions to trillions of parameters in a few years~\cite{MLSize22_Villalobos}. For example, OpenAI’s GPT-series grew from 117 million parameters in 2018 to over a trillion in GPT-4~\cite{SurveyLLM25_Zhao,GPT4Leak23}, driving unprecedented demand for memory and compute capacity~\cite{ML25_Lin,LLM20_Brown,Palm23_Chowdhery,OPT22_Susan,Touvron2023LLaMAOA}.

To meet these demands, both training and inference are distributed across multiple GPUs. This distribution is enabled through a variety of parallelization strategies—including tensor parallelism, model parallelism, data parallelism, fully sharded data parallelism (FSDP), and ZeRO~\cite{MP12_Dean,ParallelCNN18_Jia,ParallelismNN18_Gholami,Megatron-LM21_Narayanan,TensorFlow16_Abadi,weirdtrickparallelizingconvolutional14_krizhevsky,DemistifyDNN18_Tal,MDParallelism17_Gholami,ZeRO20_Rajbhandari,PyTorchDistributed20_Li,Megatron19_Shoeybi}. These techniques distribute model parameters and/or input data across GPU clusters~\cite{DGX17_Gawande, MetaMultiGPU17_Goyal, MGSIM18_Arxiv, FinePack23_Muthukrishnan, CommBench24_ICS, Frontier22_Oak, SummitSiera20_Stunkel, Summit18_OSTI} to enable scalable execution.

However, distributing computation in this way inherently introduces inter-GPU communication (e.g. weight updates, activation exchanges between layers, etc)~\cite{FusedCompCollKernels24_punniyamurthy,Centauri24_Chen} resulting in diverse and intensive communication patterns across the GPU cluster~\cite{DGX17_Gawande,MetaMultiGPU17_Goyal,MGSIM18_Arxiv,FinePack23_Muthukrishnan,CommBench24_ICS,Frontier22_Oak,SummitSiera20_Stunkel,Summit18_OSTI}. To support these patterns efficiently, collective operations like AllReduce, AllGather, AllToAll, implemented in high-performance libraries such as NCCL~\cite{NCCL17_Sylvain}, RCCL~\cite{RCCL25_AMD}, and oneCCL~\cite{ONECLL23_Intel}, have become the lifeblood of distributed ML, synchronizing gradients, routing data, and aggregating parameters efficiently.

As models scale~\cite{FewShotLearners20_OpenAI,TrainNLG22_Smith,UTTTransformer19_Raffel,DNN17_Krizhevsky,Bert18_Devlin,LanguageMA19_Radford,GPipe18_Huang,SGD17_Akiba}, so does the infrastructure: inference spans tens to hundreds of GPUs, while training cutting-edge models can require thousands~\cite{Interconnects20_Li,FrontierExascale23_Atchley,Fugaku22_Sato,meta2024_genai,InsightSuperComputerInterconnects24_Sensi}. This scaling follows a two-tier model~\cite{Tartan18_Li,NetCrafter25_Fatima}: vertical (intra-node) scaling via GPU interconnects such as NVLink~\cite{nvidia_nvlink_2017}, AMD Infinity Fabric Link~\cite{IF24_Schieffer}, or Huawei HCCS~\cite{HuaweiHCCS21_Liao}, and horizontal (inter-node) scaling via GPU-NIC interconnects using RDMA~\cite{DNNOverRDMA19_Xue,OneSidedRDMA13_Mitchell,AcceleratingSecureML24_Ren,understandingRDMAMicro23_kong,SurveyRdmaNIC24_Hu}. While intra-node links offer terabits-per-second bandwidth and direct load/store semantics~\cite{PascalGPUandNVlink17_Foley,Comscribe20_Akhtar}, inter-node communication remains a bottleneck due to slower NIC-mediated bandwidth~\cite{FuseLink25_Ren,NetCrafter25_Fatima}.

To bridge this gap, new scale-up fabrics such as NVIDIA’s NVLink network~\cite{NVLinkNetworkSwitch22_NVIDIA} and the recently ratified UALink 200G 1.0 standard~\cite{UltraEthernetUALink25_Arsid,UALinkCollab25,morgan2025ualink,UALinkSpecs25_Dave} introduce high-bandwidth, memory-semantic interconnects that allow accelerators to directly load, store, and perform atomic operations across nodes. These fabrics treat entire multi-GPU pods as unified devices, enabling full-speed all-to-all communication at pod scale.

However, such inter-node fabrics introduce a new memory abstraction: the Network Physical Address (NPA), which represents a location outside the local OS domain~\cite{morgan2025ualink,NVLinkNetworkSwitch22_NVIDIA}. To complete a remote access, the destination must translate the NPA into a System Physical Address (SPA) before servicing the request. We call this process \rt, denoting NPA-to-SPA translation at the target node. Both UALink and NVLink acknowledge \rt and propose dedicated hardware modules such as Link MMUs and Link TLBs~\cite{morgan2025ualink,NVLinkNetworkSwitch22_NVIDIA}, yet they provide little detail on their design or performance.

This lack of understanding motivates our work. 
Prior research on address translation has predominantly focused on virtual-to-physical translation at the source (CPU/GPU), with well-studied optimizations such as larger and more associative TLBs, multi-level hierarchies~\cite{Shared11_Bhatt,Borg92_TLB,cox17_MultiPageSize,LargeMem15_Karakostas,HybridTLB17_Park,pham14_IncreasingTLBReach,pham12_COLT,Synergetic10_Srikantaiah,Yan19_TransRan,Ryoo17_POM,SpecTLB11_Barr,Papadopoulou15_PredctionBasedTLB}, and prefetching techniques~\cite{vpim23_Fatima,Inter10_Bhattacharjee,TLBpre02_Kandiraju,Saulsbury00_TLBPreloading,TTP17_Bhatt} to reduce translation overhead. However, these approaches exclusively target the initiator of memory accesses leaving the destination-side translation problem largely unexplored. By shifting the translation locus to the memory target, \rt creates new bottlenecks that directly impact collective communication performance.

In this paper, we present the first systematic study of \rt in large-scale, multi-GPU, multi-node systems. We extend \textit{ASTRA-sim}~\cite{AstraSim23_Won} with Omnet++~\cite{Omnetpp08_Varga} as the network backend to model Link MMUs and Link TLBs, enabling detailed evaluation of their impact on collective communication workloads. Our analysis shows that \rt introduces non-trivial translation overheads, particularly under cold TLB states in smaller collectives—a common case in inference workloads—directly degrading communication throughput. We find that performance is primarily dictated by the translation working set of a collective, which scales with the number of participating GPUs. Increasing TLB capacity yields diminishing returns once it is sufficient to cover this working set. This explains why simply scaling TLB size beyond this threshold provides limited benefit. Instead, a modestly sized TLB can be highly effective if cold misses are mitigated through techniques such as software prefetching or fused pre-translation kernel.


To summarize, the main contributions of this paper are:




\begin{itemize}
\item We conduct the first in-depth performance analysis of \rt across varying collective input sizes and GPU counts, showing that it can introduce significant overheads that degrade end-to-end collective performance.

\item We show that performance is most impacted during system warm-up, where cold misses in high-latency translation modules dominate. Although caches improve over time, the penalty of cold misses remains substantial, especially for small, latency bound collectives.

\item To mitigate these overheads, we propose two directions for optimization: (1) fused pre-translation kernels to proactively warm Link TLBs during compute phases, and (2) software prefetching of TLB entries to hide translation latency.  

\item We extend \textit{ASTRA-sim}~\cite{AstraSim23_Won} with Omnet++~\cite{Omnetpp08_Varga} as the network backend, implementing detailed models of \rt modules. This enables accurate simulation and characterization of their impact on inter-GPU collective communication.

    
    


\end{itemize}

%% file: sec/2_background.tex
\section{Background}
\subsection{Multi-Node Multi-GPU Systems}
Modern AI~\cite{SurveyLLM25_Zhao,GPT4Leak23} workloads increasingly outgrow the capacity of a single accelerator, motivating the use of multi-GPU systems. A multi-GPU system~\cite{CommBench24_ICS} refers to multiple accelerators within a single physical host. These GPUs share the same CPU complex, operating system image, and often communicate through a low-latency interconnect fabric such as PCIe switches, AMD Infinity Fabric Link~\cite{IF24_Schieffer}, NVLink~\cite{nvidia_nvlink_2017}, or NVSwitch~\cite{NVLinkNVSwitch22_Patrick}. Such configurations enable efficient data sharing with relatively low communication overhead, but are typically limited to 4–16 GPUs per node due to physical constraints in power delivery, board space, and cooling.

Scaling further requires multi-node multi-GPU systems, in which multiple GPU-equipped hosts are connected via network fabrics such as InfiniBand or Ethernet RDMA~\cite{DNNOverRDMA19_Xue,OneSidedRDMA13_Mitchell,AcceleratingSecureML24_Ren,understandingRDMAMicro23_kong,SurveyRdmaNIC24_Hu}. These clusters allow hundreds or thousands of accelerators to train a model in parallel, but at the cost of higher communication latency, weaker memory-sharing semantics, and complex topology-aware scheduling. In these systems, collective communication libraries (e.g., NCCL, RCCL, or MPI) are essential to coordinate communication across nodes, but software overheads and network bottlenecks often become limiting factors, highlighting the need for an open, low-latency accelerator-to-accelerator fabric. Proprietary solutions like NVIDIA’s NVLink/NVSwitch~\cite{NVLinkNVSwitch22_Patrick,NVIDIADGXGB20024_Patrick} offer high performance but are platform-specific, whereas PCIe and RDMA-based solutions provide openness at the expense of latency and efficiency~\cite{RDMAOverhead_2021,BenchmarkingGPUDirectRDMA14_Rossetti,unat2024landscape}.

UALink addresses this gap. Ratified in 2025 by a broad industry consortium~\cite{UALinkEvaluationCopy25_UALink,UALinkCollab25,morgan2025ualink}, it defines an open accelerator-to-accelerator interconnect and switching standard. UALink supports direct load/store and atomic operations between accelerators without requiring host mediation. Dedicated UALink Switches (ULS)~\cite{UALinkEvaluationCopy25_UALink} enable the formation of “pods” containing up to 1,024 accelerators. Communication within a pod occurs over the low-latency UALink fabric, while Ethernet or InfiniBand is used only for inter-pod transfers. Critically, the specification is governed openly, allowing multiple vendors to participate and reducing the risk of proprietary lock-in. 

\subsection{Communication through UALink}
In an UALink-connected system, an accelerator can communicate with another accelerator either via a direct UALink link or through a UALink switch. Communication within the same system node is referred to as in-domain or intra-node communication, while communication across different nodes is called cross-domain or inter-node communication~\cite{UALinkEvaluationCopy25_UALink}.

\begin{figure}[t]
    \centering    
    \includegraphics[width=\linewidth]{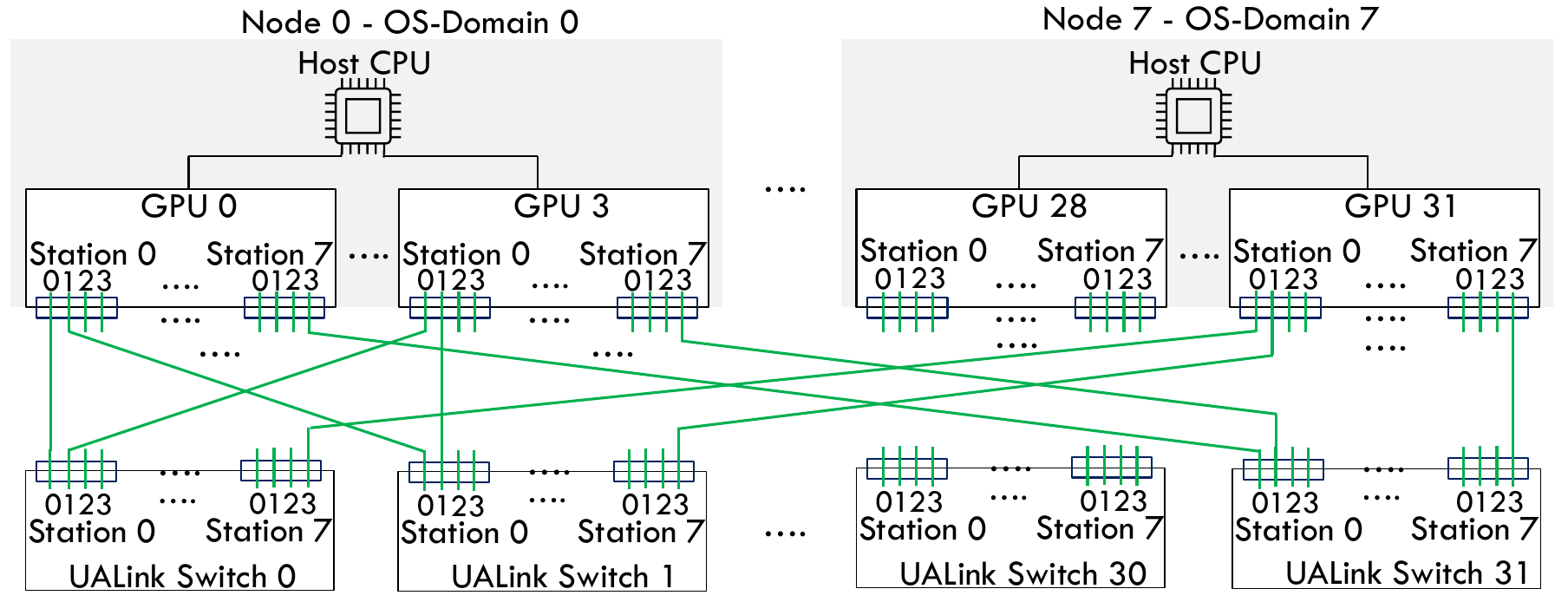}
    \caption{A multi-node multi-gpu system connected over a UALink network (For clarity, certain of the UALink Links have been shown while the other links are omitted).}
    \Description{}
    \vspace*{-\baselineskip}
 \label{fig:ualink}
\end{figure}

UALink Switches enable a direct load/store access model for a scale-up accelerator pod with up to 1,024 accelerators. Each port on an accelerator interconnects with only one port on every other accelerator in the pod. The specification supports a maximum data rate of 200 GT/s per lane and a maximum link width of 4 lanes. A UALink Station (or simply Station) comprises four lanes and may be bifurcated into one x4, two x2, or four x1 UALink links. Links connect ports on accelerators to ports on UALink switches. The maximum bandwidth for each station is 800 Gbps. All stations in a pod (both on switches and accelerators) must follow the same bifurcation pattern, and ports on each accelerator are identically numbered. Requests and responses are routed from a source port on the source station, through the switch, to the target station, and finally to the target port. In a pod, a physical switch has at least as many ports as there are accelerators. For example, a 32-accelerator pod uses 32 switches with 32 x1 UALink links each, allowing every accelerator to connect to each switch via a dedicated port (see Figure~\ref{fig:ualink}).

\subsection{\rt}
\begin{figure}[ht]
    \centering    
    \includegraphics[width=\linewidth]{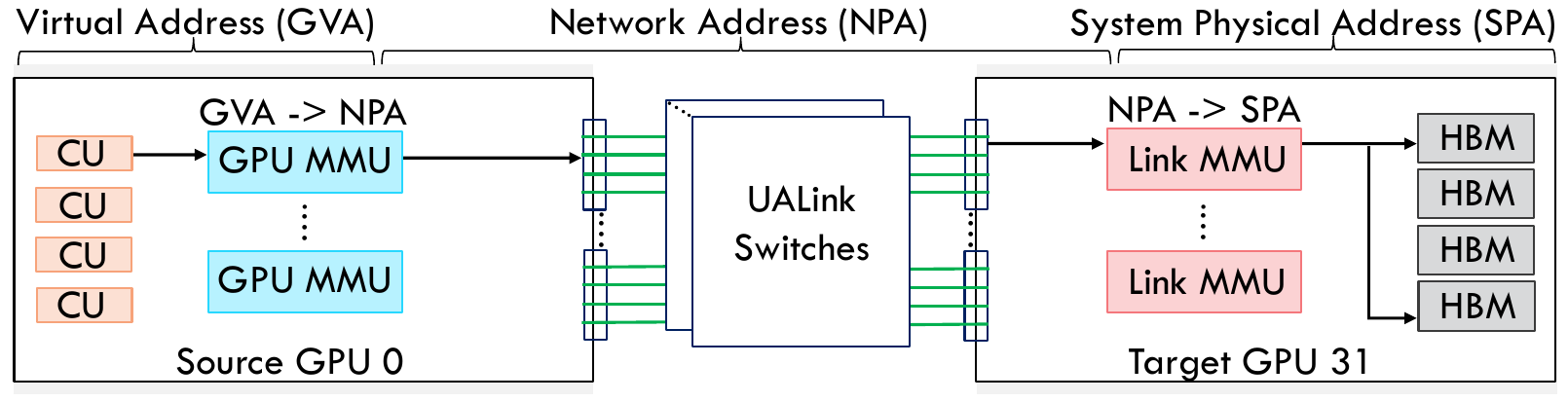}
    \caption{\rt of a Network Physical Address (NPA) to a System Physical Address (SPA) at the target GPU for inter-node accesses.}
    \Description{}
    \vspace*{-\baselineskip}
 \label{fig:ualink_reverse_translation}
\end{figure}

Historically, GPUs within a system used System Physical Addresses (SPAs) to reference memory. With inter-node communication over UALink, a new addressing model—the Network Physical Address (NPA)—is introduced, as illustrated in Figure~\ref{fig:ualink_reverse_translation}. When a GPU initiates a memory access, its virtual address is translated by the GPU's Memory Management Unit (MMU) into either an NPA, for accessing memory on a remote node (i.e., a different OS domain), or an SPA, for accesses within the same OS domain. SPA remains the standard for local memory accesses. For inter-node accesses, the target GPU must convert the incoming NPA back to the corresponding SPA before performing the memory operation. This \rt (NPA → SPA) is handled by the Link MMU on the target GPU, which traverses a series of translation modules to correctly resolve the network-level address to a local system-physical address (Figure~\ref{fig:ualink_reverse_translation}).

\subsection{Baseline System}

\begin{figure}[t]
    \centering    
    \includegraphics[width=\linewidth]{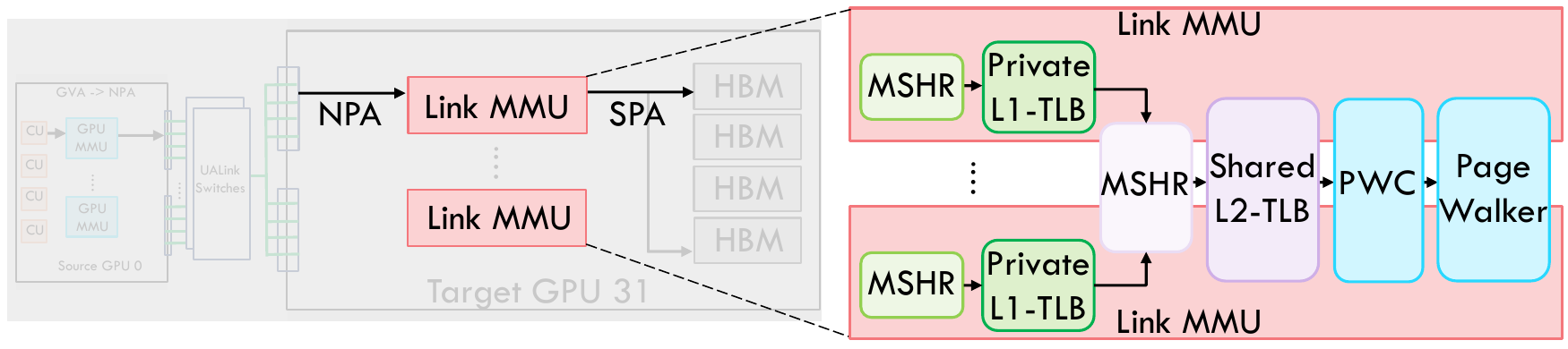}
    \caption{Our baseline \rt hierarchy for performing \rt at the Target GPU node.}
    \Description{}
    \vspace*{-\baselineskip}
 \label{fig:ualink_reverse__translation_baseline}
\end{figure}

While the UALink specification~\cite{UALinkEvaluationCopy25_UALink} describes the \rt and the Link MMU at the target GPU that performs translation, it provides little detail on the internal translation modules themselves. Moreover, since this is the first work to characterize this type of reverse address translation at the target node, no prior studies exist for use as a baseline. To address this, we adopt a structural approach similar to prior GPU IOMMU designs~\cite{TransFW23_Li} for our evaluation and impact analysis. Figure~\ref{fig:ualink_reverse__translation_baseline} illustrates the translation hierarchy used throughout this paper. Each UALink station has a private L1 Link TLB. Accesses that miss in the L1 TLB are forwarded to a shared L2 Link TLB, which services traffic from all UALink stations. Misses at the L2 level are sent to page walk caches, and if still unresolved, to the page table walker (PTW), which is also shared across all UALink traffic at the target GPU. The Link TLB hierarchy employs a mostly-inclusive policy~\cite{AdvancedConceptAT17_Bhattacharjee}: when the PTW is triggered due to an L2 miss, the translation is populated into both the L1 and L2 TLBs. Conversely, evictions from a lower-level TLB do not require invalidation in higher-level TLBs.

\subsection{Collectives In Distributed ML Models}
Large ML models use parallelism strategies including tensor parallelism, model parallelism, data parallelism, fully sharded data parallelism (FSDP), and ZeRO~\cite{MP12_Dean,ParallelCNN18_Jia,ParallelismNN18_Gholami,Megatron-LM21_Narayanan,TensorFlow16_Abadi,weirdtrickparallelizingconvolutional14_krizhevsky,DemistifyDNN18_Tal,MDParallelism17_Gholami,ZeRO20_Rajbhandari,PyTorchDistributed20_Li,Megatron19_Shoeybi} to avoid data duplication across distributed nodes. However, such strategies result in additional collective communication which is used to coordinate and share data during training (or inference) to exploit parallelism across the multiple nodes using various communication algorithms~\cite{msccl23_Cowan,OptCollCommMPICH05_Thakur}. 

AlltoAll~\cite{AlgorithmAllToAll94_Bruck} communication is a fundamental collective operation in distributed ML workloads. It is widely used in model parallelism, for exchanging activations and gradients across layers, and in Mixture-of-Experts (MoE) architectures, for routing tokens to different expert sub-layers. In MoEs, All-to-All collectives occur twice per layer: once for dispatching inputs to the experts and again for gathering outputs, forming a critical path that can account for a significant portion of execution time~\cite{Tutel23_Hwang}. Given its prevalence and the performance impact observed in modern large-scale models, we focus our evaluation on All-to-All operations, which represent a common and performance-critical communication pattern in distributed ML.

In the following sections, we outline our methodology, which forms the basis for our characterization study of the performance overheads of \rt at the target node on All-to-All collective communication workloads.

%% file: sec/3_methodology.tex
\section{Methodology}
We evaluate the most common and high traffic collective: All-to-All that is generated from the MSCCLang frameworks~\cite{msccl23_Cowan}. To evaluate, we use the ASTRA-sim simulator~\cite{AstraSim23_Won}. AstraSim runs isolated collective communication kernels described in XML or JSON formats, often generated by synthesizers, and integrates workload to traffic generation with a variety of network backends. Specifically, we use the Omnet++~\cite{Omnetpp08_Varga} framework as our network simulation backend. Omnet++ is a discrete-event, component-based simulation framework that enables detailed modeling of packet-level network behavior, including link contention, queueing, congestion control, flow-level crediting, and routing protocols. Its modular architecture allows integration of custom network models, such as the UALink interconnect used in our study. In our setup, GPUs are connected via a railed, single-level Clos backbone network of high-radix Ultra Accelerator Link (UAL) switches. While UALink parameters are evolving, our simulation uses the most up-to-date bandwidth and latency values, as listed in Table\ref{tab:gpu-config}. Detailed baseline system specifications are presented in Table~\ref{tab:gpu-config}. We evaluate a page size of 2MB.

The workloads were generated with MSCCLang example scripts for the all-pairs/direct algorithm~\cite{msccl-tools}. All schedules for MSCCLang are two-sided and use remote store instructions. In all measurements, the “size” of the collective is the larger size of a single GPU’s input or output buffer.
In the all-pairs algorithm, at each GPU source, a unique WG transmits a chunk of data to each destination. Since this work is focused on communication collectives that do not have data reuse in caches, we assume memory accesses missing in all cache levels and a constant 120ns latency for a request to traverse the cache hierarchy from a CU to the NoC fabric.
\input{tab/baseline}

%% file: tab/baseline.tex
\begin{table}[h]
\centering
\caption{Simulation Setup }
\begin{tabular}{|l|l|}
\hline
\multicolumn{2}{|c|}{\textbf{System}} \\ \hline
\# GPUs & 8, 16, 32, 64 (4 GPUs/node)\\ \hline
 Inter-GPU Link & UALink single-level Clos 
 network~\cite{UALinkSpecs25_Dave}  \\ \hline
 Local Data Fabric & 120ns latency \\ \hline

\multicolumn{2}{|c|}{\textbf{Per GPU Config}} \\ \hline
Dies & 8x compute, 2x I/O~\cite{AMDMi350x_25} \\ \hline
Compute Unit & 2.2 GHz, 256 per GPU~\cite{AMDMi350x_25}  \\ \hline
HBM & 150ns access latency~\cite{OuterSPACE18_Pal} \\ \hline

\multicolumn{2}{|c|}{\textbf{Reverse Translation Config~\cite{morgan2025ualink,NVLinkNetworkSwitch22_NVIDIA}}} \\ \hline
L1 Link TLB & 32-entry ~\cite{TransFW23_Li}, fully-assoc, 50 ns hit lat~\cite{DissertingGPU18_Jia}, \\& private/UALink Station, 256-entry MSHR
\\ \hline
L2 Link TLB & 512 entry~\cite{TransFW23_Li}, 2-way set assoc, 100 ns hit \\& lat~\cite{DissertingGPU18_Jia}, LRU replacement policy, shared \\& across UALink stations per GPU\\ \hline
Link MMU~\cite{morgan2025ualink} 
& 5-level page table with page walk cache \\ & (16,32,64,128 entries~\cite{TransFW23_Li}, 2-way, 50ns latency) \\ & , shared walker, 100 parallel PTWs \\ \hline

\multicolumn{2}{|c|}{\textbf{Inter-GPU UALink Configuration}} \\ \hline
UALink Switch & Single level clos, 300ns latency ~\cite{Ualinkswitchlatency25_Mann, UALinkEvaluationCopy25_UALink} \\ \hline

UALink Station & 16 per GPU, 4 lanes per station (combined \\ & as 1 x4 port or link), 200Gbps effective \\& BW/lane~\cite{UALinkEvaluationCopy25_UALink} \\ \hline


UALink Link & 800 Gbps cumulative bandwidth, 300 ns \\ & die-to-die latency~\cite{NVLinkLatency25_Compute} \\ \hline
\end{tabular}
\label{tab:gpu-config}
\end{table}

%% file: sec/4_key_ideas.tex
\section{Evaluation and Analysis}
\subsection{\rt Overheads}
\begin{figure}[ht]
    \centering    
    \includegraphics[width=\linewidth]{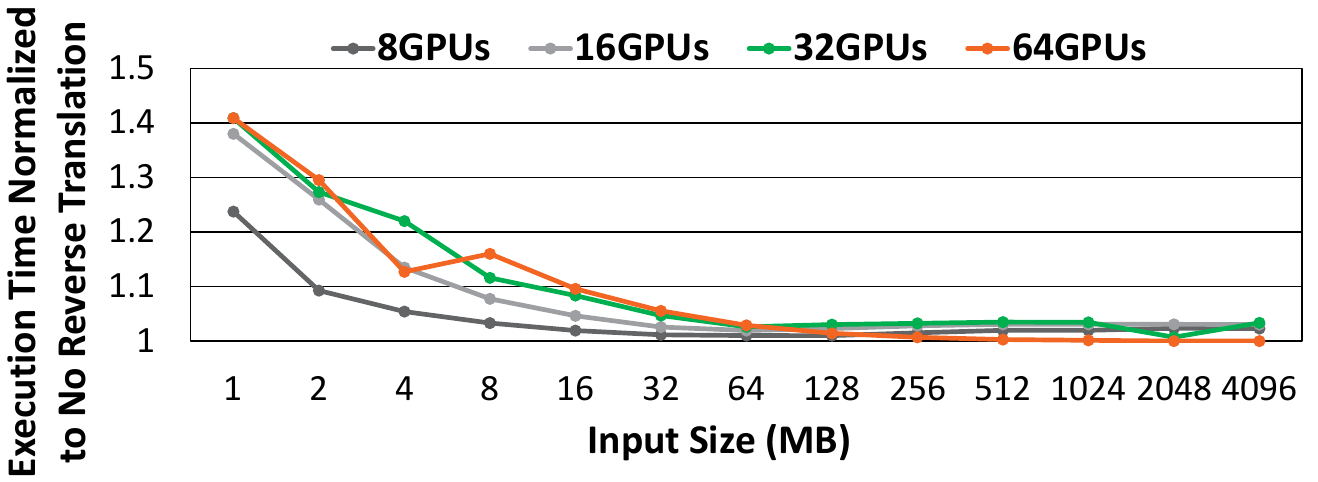}
    \caption{Performance overhead of \rt, normalized to an ideal configuration with zero \rt overhead, evaluated on systems with eight and up to 64 GPUs with AlltoAll collective sizes ranging from 1 MB to 4 GB.}
    \Description{}
    \vspace*{-\baselineskip}
 \label{fig:overhead}
\end{figure}

To evaluate the impact of \rt on multi-node, multi-GPU performance, we compare our baseline configuration against an \textit{ideal} setup, where \rt introduces zero overhead. While the ideal case is not practically achievable, it serves as an upper-bound for potential optimization. Figure~\ref{fig:overhead} presents performance degradation across GPU pod sizes ranging from 8 to 64 GPUs and collective communication sizes from 1MB to 4GB.

We observe that small collectives (1MB) experience up to 1.4$\times$ performance degradation, while larger collectives (16MB) incur only around 1.1$\times$ overhead. This trend indicates that \rt overhead is most pronounced for small collective sizes, where each request is more likely to encounter cold page table walks, and gradually diminishes as collective size increases. The results highlight a critical interplay between request volume and translation latency: larger collectives naturally amortize the cost of translation across many requests, while smaller collectives are disproportionately affected by individual high-latency translations.

\subsection{Quantifying Per-Request Latency}
\begin{figure}[ht]
    \centering    
    \includegraphics[width=\linewidth]{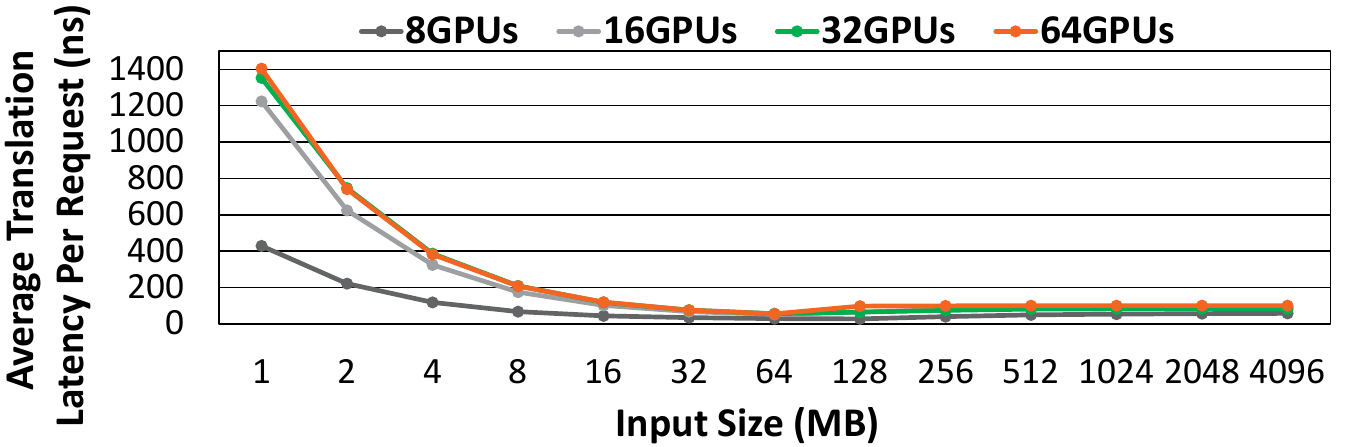}
    \caption{Average \rt latency per request, evaluated on systems with eight and up to 64 GPUs with AlltoAll collective sizes ranging from 1 MB to 4 GB.}
    \Description{}
    \vspace*{-\baselineskip}
 \label{fig:avgtranslation_latency}
\end{figure}

To understand the observed performance trends, we analyze the average \rt latency per request for the same sweep of GPU pod sizes and collective sizes (Figure~\ref{fig:avgtranslation_latency}). The results confirm that small collectives experience high per-request \rt latency due to cold TLBs and page table walks, whereas larger collectives benefit from warmed caches and TLB entries, which reduce latency significantly.

\begin{figure}[ht]
    \centering    
    \includegraphics[width=\linewidth]{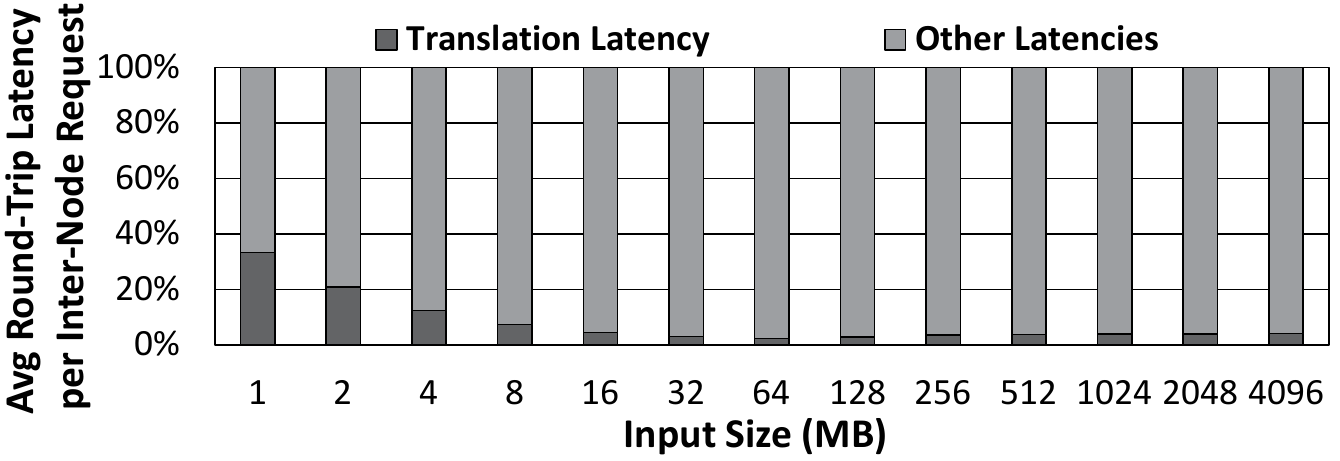}
    \caption{Fraction of the round trip latency per request spent in performing \rt and other latencies evaluated for 16-GPU configuration and varying AlltoAll collective size.}
    \Description{}
    \vspace*{-\baselineskip}
 \label{fig:avg_roundtrip_latency}
\end{figure}

Figure~\ref{fig:avg_roundtrip_latency} provides a stacked breakdown of the round-trip latency per request for a 16-GPU system. For 1MB collectives, up to \textasciitilde30\% of total request latency is spent performing \rt. This fraction steadily decreases with increasing collective size, reinforcing the observation that \rt overhead is amortized more effectively in larger collectives. These figures illustrate not only the absolute latency contributions of \rt but also how its relative significance diminishes with larger data movement.

\subsection{Hierarchical Translation Scenarios}
\begin{figure}[ht]
    \centering    
    \includegraphics[width=\linewidth]{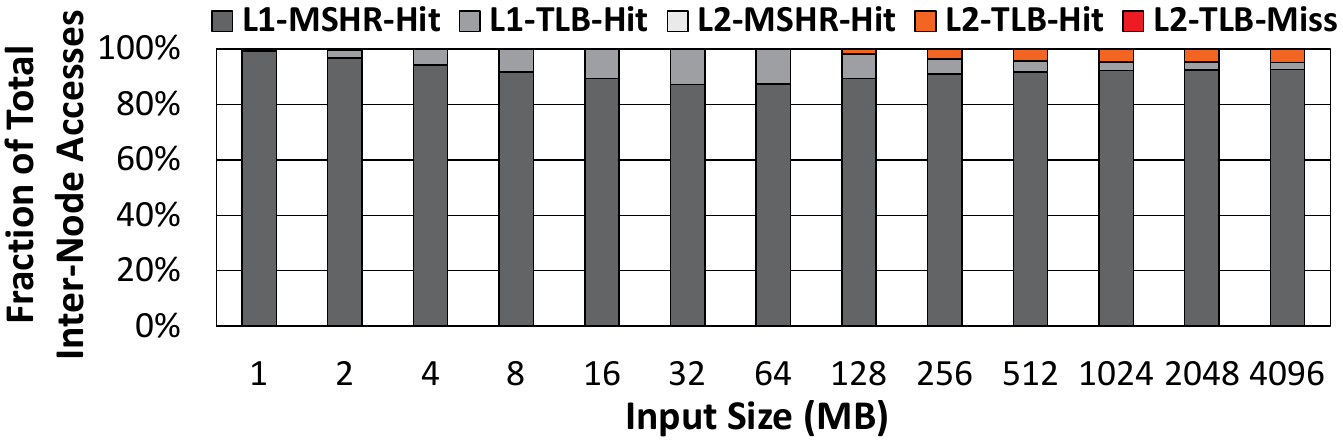}
    \caption{Stacked breakdown of \rt hits and misses at target GPU translation modules for inter-node requests in a 16-GPU system under varying AlltoAll sizes.}    
    \Description{}
    \vspace*{-\baselineskip}
 \label{fig:fraction_internode}
\end{figure}

Next, we examine the sources of \rt overhead by investigating the translation hierarchy at the target GPU node. Figure~\ref{fig:fraction_internode} shows the breakdown of all inter-node requests originating from a source GPU in a 16-GPU system. Over 90\% of requests hit the L1-MSHR; however, this metric alone is insufficient to predict latency, as requests may still stall due to pending page walks at lower levels of the hierarchy.

\begin{figure}[ht]
    \centering    
    \includegraphics[width=\linewidth]{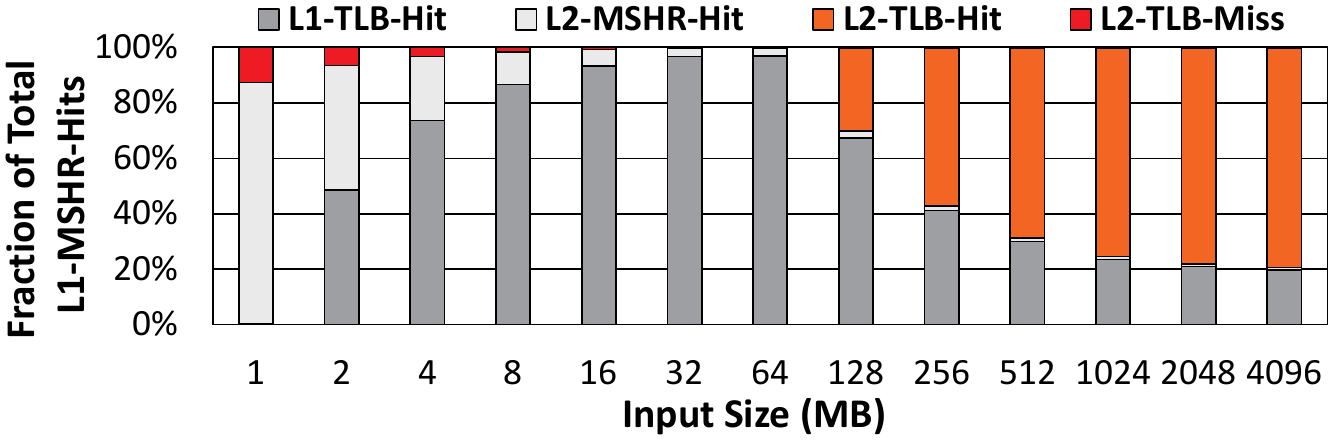}
    \caption{Stacked breakdown of L1-MSHR hit-under-miss and miss scenarios at target GPU translation modules for all inter-node requests from a source GPU in a 16-GPU system across varying AlltoAll sizes.}
    \Description{}
    \vspace*{-\baselineskip}
 \label{fig:breakdown_fraction_internode}
\end{figure}

Figure~\ref{fig:breakdown_fraction_internode} further decomposes L1-MSHR hits into hits-under-miss and misses at subsequent hierarchy levels. For 1MB collectives, L2-TLB misses and L2-TLB-hit-under-miss scenarios dominate, reflecting cold page walks at the lowest hierarchy levels. As input size increases (2–64MB), L1-TLB hits gradually dominate, indicating that initial page walks warm the translation hierarchy. At 64MB, the 32-entry L1-TLB reaches capacity; however, overall performance remains stable because L2-TLB hits compensate for the reduction in L1-TLB hits. This shows that hierarchical caching effectively mitigates performance degradation even when lower-level caches are saturated..

\subsection{Per request \rt Latency Patterns}
\begin{figure}[ht]
    \centering    
    \includegraphics[width=\linewidth]{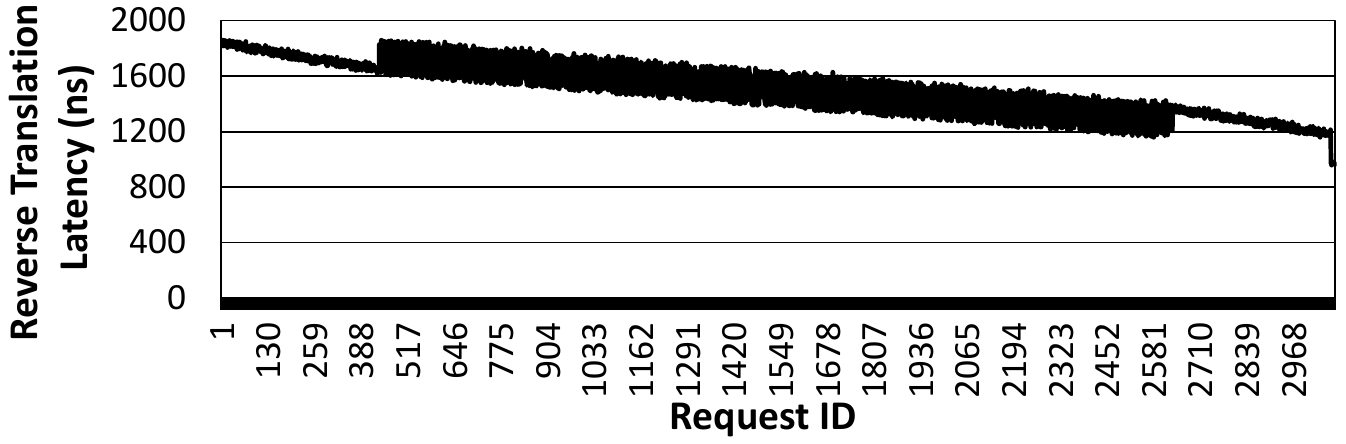}
    \caption{\rt latency per request plotted for all request trace originating from source GPU node for a 1MB input size and 16 GPU configuration.}
    \Description{}
    \vspace*{-\baselineskip}
 \label{fig:per_request_graph_1MB}
\end{figure}

\begin{figure}[ht]
    \centering    
    \includegraphics[width=\linewidth]{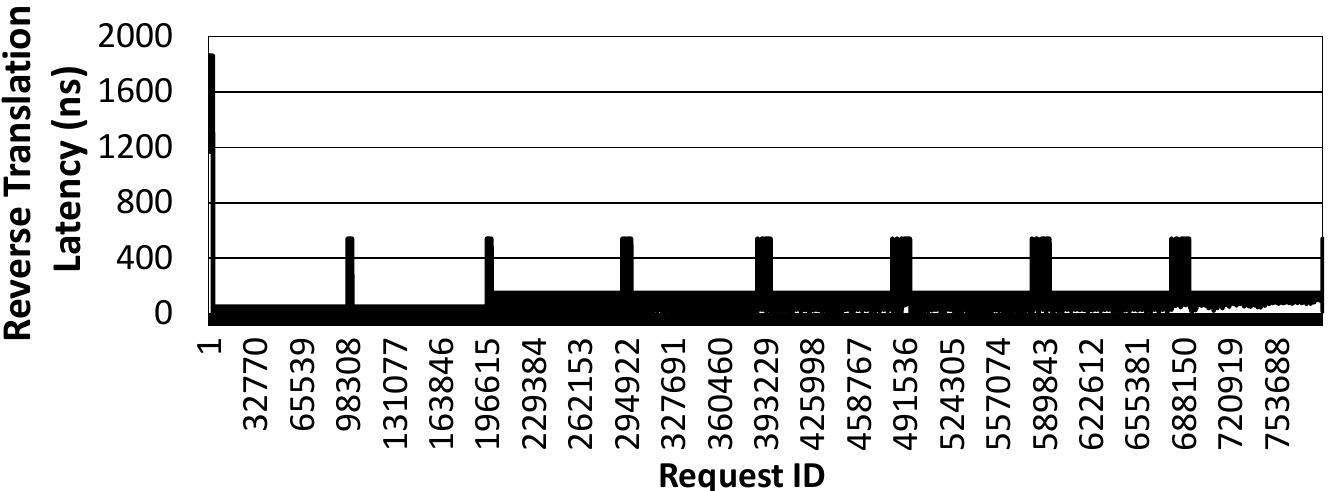}
    \caption{\rt latency per request for all request trace originating from source GPU node for a 256MB input size and 16-GPU configuration.}
    \Description{}
    \vspace*{-\baselineskip}
 \label{fig:per_request_graph}
\end{figure}

We next analyze per-request \rt latency for small (1MB) and medium (256MB) collectives, as shown in Figures~\ref{fig:per_request_graph_1MB} and~\ref{fig:per_request_graph}.

For 1MB collectives, all requests originating from the source GPU encounter high \rt latency due to cold page table walks across destination GPUs. These initial misses generate the severe performance degradation observed in Figure~\ref{fig:overhead}, as each request must traverse the full translation hierarchy. At this small input size, TLBs and page walk caches are largely cold, so latency is dominated by the cost of resolving new pages.

In contrast, the 256MB collective exhibits multiple latency spikes corresponding to requests accessing cold pages across destination GPUs. The first spike represents initial cold misses across all destinations, where page table walks occur for the first time. After these entries are populated in the destination TLBs, subsequent accesses largely hit warmed TLB entries, significantly reducing \rt latency. Additional smaller spikes occur when request offsets exceed page boundaries: these accesses partially hit in the page walk caches, which avoid full table walks and mitigate latency.

This spike pattern reflects the streaming access pattern of custom collectives: each page is accessed sequentially with strides, exploiting spatial locality within the page. Once the stride moves to a new page, the old page is rarely reused, demonstrating minimal temporal locality. Importantly, this means that at any point, the destination GPU sees at most 1$\times$(number of GPUs) pages simultaneously, as each participating GPU contributes only one active page at a time. This insight explains why L2-TLB overprovisioning does not improve performance for these workloads.

Overall, these observations highlight that \rt overhead is dominated by initial cold misses. Hierarchical caching—through TLBs and page walk caches—effectively amortizes translation costs across subsequent requests. Furthermore, the combination of stride-based accesses and streaming behavior ensures that once page entries are warmed, repeated accesses are rare, stabilizing per-request latency for larger collectives.

\subsection{Impact of L2-TLB Sizes}
The preceding analysis indicates that custom collectives access each page in a mostly streaming manner: requests stride through the data within a page, exploiting spatial locality, but once the stride moves to a new page, the previously accessed page is rarely revisited. As a result, temporal locality is minimal. Consequently, the L2-TLB only needs to accommodate the number of pages simultaneously accessed across all participating GPUs; beyond that, increasing its size provides little to no performance benefit.

\begin{figure}[ht]
    \centering    
    \includegraphics[width=\linewidth]{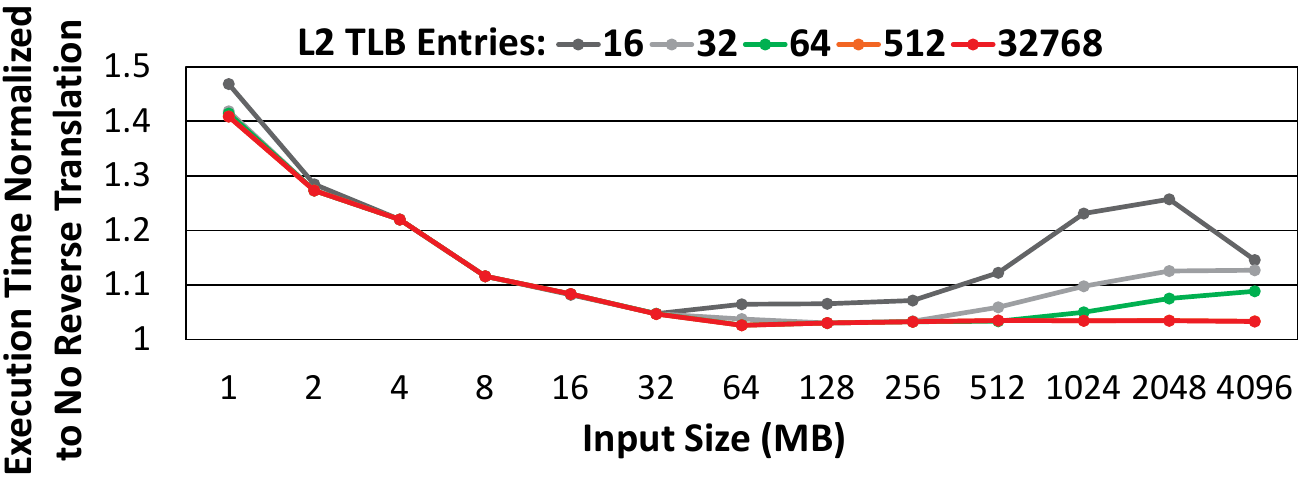}
    \caption{Performance overhead of \rt, normalized to an ideal configuration with zero \rt overhead, evaluated on 32 GPUs with AlltoAll collective sizes ranging from 1 MB to 4 GB with varying L2-TLB sizes.}
    \Description{}
    \vspace*{-\baselineskip}
 \label{fig:tlb_sensitivity}
\end{figure}

To validate this, we vary L2-TLB sizes from 16 to 32, 64, 512, and 32,768 entries and measure performance for a 16MB input size on 32 GPU configuration (Figure~\ref{fig:tlb_sensitivity}). Even with a small L2-TLB of 32 entries—equal to the number of GPUs accessing a page simultaneously—performance remains stable. Larger L2-TLB sizes do not improve performance further, confirming that over-provisioning L2-TLBs for ML collective workloads is unnecessary. This insight suggests that modest L2-TLB capacities suffice to sustain high-performance collective communication, reducing hardware cost and complexity without impacting runtime efficiency.

%% file: sec/5_summary.tex
\section{Summary}
Based on our analysis of the most traffic-intensive AlltoAll collectives commonly used in ML workloads, we identify two key observations: 1) Cold misses dominate small-collective performance: For latency-sensitive collectives, especially small ones (e.g., 1MB), cold TLB misses contribute disproportionately to total latency. Our measurements show that \rt overhead can degrade performance by up to 1.4$\times$ for 1MB collectives, and this impact could be even higher for smaller collectives. This effect arises because each request must traverse the full page translation hierarchy, including potentially multiple page table walks across destination GPUs. These cold misses dominate the critical path for latency-sensitive workloads, making them a primary target for optimization. 2) Minimal temporal locality reduces L2-TLB requirements: Due to the streaming access patterns of custom collectives—where requests stride sequentially across pages, rarely revisiting previous pages—the effective temporal locality is minimal. Consequently, L2-TLB sizes exceeding the number of participating GPUs provide negligible performance improvement. Modest L2-TLB capacities that accommodate at least one active page per GPU suffice to sustain high throughput, reducing the need for costly over-provisioning.

These findings have direct implications for ML workloads. While training workloads often process large batches that saturate bandwidth, amortizing point-to-point network latency, inference workloads are increasingly latency-sensitive, often operating on small batches. Inference can account for the majority of computational resources and costs in large-scale deployments at major tech firms such as Meta, Amazon, and and Google~\cite{CarbonFootprintML22_patterson,SustainableAi22_Wu}. Network latency in inference can consume up to 20\% of total runtime for modern LLMs~\cite{InferenceEconomiesLLM25_Erdil,TokenWeave25_Gond}, making \rt overhead in small collectives a critical performance bottleneck.

\section{Observations and Opportunities}
Based on these insights, we propose two avenues for mitigating \rt overhead in latency-sensitive collectives: 1) Pre-Translation and Fused Kernels:
Integrate pre-translation requests directly into computation kernels, allowing page table entries to be fetched ahead of the corresponding data access. By overlapping pre-translation with computation, the effective latency for subsequent requests can be hidden. This approach is particularly effective for ML based collectives where communication patterns are predictable and stride-based. 2) Software-Driven TLB Prefetching: Implement software-guided prefetching mechanisms that predictively populate TLBs with pages likely to be accessed next. Prefetching strategies could leverage static memory layout knowledge of input, output, and scratch buffers, or dynamic profiling to detect repetitive access patterns across GPUs. This reduces cold-miss penalties without increasing hardware TLB sizes.

In summary, our analysis reveals that small, latency-sensitive collectives suffer heavily from cold TLB misses, whereas L2-TLB overprovisioning provides little benefit due to minimal temporal locality. Optimizing \rt performance in small collectives is therefore crucial, especially for inference-heavy ML workloads. The technical opportunities outlined—ranging from pre-translation fused kernels to predictive page table replication—provide concrete directions to reduce \rt overhead, improve end-to-end collective performance, and unlock more efficient multi-GPU deployments for modern AI workloads.

%% file: sec/7_conclusion.tex
\section{Conclusion}
We analyzed \rt overheads in multi-node, multi-GPU systems, focusing on AlltoAll collectives for ML workloads. Our study shows that cold TLB misses dominate latency for small collectives, causing up to 1.4$\times$ performance degradation, whereas larger collectives benefit from warmed caches and L2-TLB capacity is rarely a bottleneck. To address this, we propose two avenues for future exploration: (i) pre-translation fused kernels, which overlap address translation with computation by initiating translations before communication kernels start, and (ii) software-guided TLB prefetching, which proactively populates predictive page entries. These approaches aim to mitigate translation latency for small, latency-sensitive collectives in inference workloads. Our analysis highlights the critical importance of \rt optimization for small collectives, where both translation and network latency are significant performance factors, and motivates future work on latency-hiding techniques to improve scalability in modern ML systems.